\documentclass[pre,showpacs,twocolumn]{revtex4}
\usepackage{epsfig}
\newcommand{\av}[1]{\left<#1\right>}

\begin{document}
\title{Basins of attraction on random topography}
\author{Norbert Schorghofer}
\author{Daniel H. Rothman}
\affiliation{Department of Earth, Atmospheric, and Planetary Sciences,\\
Massachusetts Institute of Technology,
Cambridge, MA 02139}
\date{\today}

\begin{abstract}
We investigate the consequences of fluid flowing on a continuous surface upon the geometric and statistical distribution of the flow.  We find that the ability of a surface to collect water by its mere geometrical shape is proportional to the curvature of the contour line divided by the local slope.  Consequently, rivers tend to lie in locations of high curvature and flat slopes.  Gaussian surfaces are introduced as a model of random topography.  For Gaussian surfaces the relation between convergence and slope is obtained analytically.  The convergence of flow lines correlates positively with drainage area, so that lower slopes are associated with larger basins.  As a consequence, we explain the observed relation between the local slope of a landscape and the area of the drainage basin geometrically.  To some extent, the slope-area relation comes about not because of fluvial erosion of the landscape, but because of the way rivers choose their path.  Our results are supported by numerically generated surfaces as well as by real landscapes.
\end{abstract}


\maketitle

%\noindent
%Keywords: statistical topography, slope-area relation, geomorphology, topographic convergence

\section{Introduction}

Aside from their natural beauty \cite{blu92} the morphological features of landscapes also bear the signatures of tectonics and past and present climates.  Hence it is important to understand their form and origin \cite{sch61}.  The large number and variety of geological processes acting in the formation of landscapes makes this a problem of formidable complexity.  Numerous hypothetical models have been proposed to help understand phenomena common to many river systems on Earth \cite{picturebook,OCN_review,dodds00}.  In this paper, we approach to problem by studying fluid flow on continuous surfaces.  This straightforward approach leads to insights about the origin of observed statistical features in real landscapes. 

Much attention has been devoted to the so-called slope-area law, that presumably relates the erosion of a landscape to the flow of water on it \cite{kirkby71,fli74,will91,picturebook,gia00,hk83,hsd94,sinclair96,banavar97,somfai97}.  The area, projected onto the horizontal, that discharges its rain water into a surface element downhill is referred to as the drainage area $a$ of the surface element.  The observed correlation of local slopes with drainage area has often been attributed to erosion mechanisms \cite{picturebook}.  A popular hypothesis is to balance downward sediment mass transport with tectonic uplift (e.g. \cite{hk83,hsd94,how94a}).  Here we show that such a correlation is found on simple surfaces for purely geometric reasons.  The main results of this paper are an exact relation, eq.~(\ref{cg1}), between slope and convergence for a wide class of random landscapes, and a geometrical explanation of the slope-area relation.  The basic, intuitive reason for both results is that flatter points on a surface have a stronger tendency to accumulate water from their neighborhood.  Flow aggregates preferably on flatter regions.

The primary distinction between our approach and that of many previous investigations is that we look at the geometrical effects of a surface rather than the active role of rivers.  Landscapes can be actively formed by continual erosion, but where do rivers {\it choose} to form on a prescribed landscape?  What are their statistical properties in the absence of any physical mechanism altering the landscape?  Studying flow on prescribed surfaces allows to understand the passive role of rivers.  

A useful family of artificial landscapes are Gaussian surfaces, which are formally introduced in the text.  They form a broad, yet relatively simple class of random surfaces.  Their topography has rich implications upon the statistical properties of flow on them.  Gaussian surfaces may therefore serve as a ``null hypothesis'' that represents basic geometric effects.  To determine the traces of erosion in landscape topography, one needs to find out what is special in the statistics beyond these geometrical effects.

In Section~\ref{sec:convergence} we ask how the geometric form of a surface determines the drainage behavior, leading to a differential expression for the topographic convergence of a surface.  Section~\ref{sec:gaussian} deals with Gaussian surfaces, which serve as our model system.  In the fourth section we demonstrate that a slope-area relation is present on Gaussian surfaces and its explanation applies also to real landscapes.  In Section~\ref{sec:new} the slope-area relation is explored more quantitatively and its limitations are pointed out.  Conclusions are reiterated and discussed in the last section.

\section{Flow convergence on a surface}
\label{sec:convergence}

We consider a smooth surface described by its height $h(x,y)$ as a function of the horizontal coordinates $x$ and $y$.  Water is assumed to follow the path of steepest descent, neglecting its inertia and volume.

\begin{figure}
% GNUPLOT: LaTeX picture
\setlength{\unitlength}{0.240900pt}
\ifx\plotpoint\undefined\newsavebox{\plotpoint}\fi
\sbox{\plotpoint}{\rule[-0.200pt]{0.400pt}{0.400pt}}%
\begin{picture}(1275,765)(0,0)
\font\gnuplot=cmr10 at 10pt
\gnuplot
\sbox{\plotpoint}{\rule[-0.200pt]{0.400pt}{0.400pt}}%
\put(566,716){\makebox(0,0)[l]{$h$}}
\put(651,581){\makebox(0,0)[l]{$h-\delta$}}
\put(246,564){\makebox(0,0)[l]{$\vec t$}}
\put(355,437){\makebox(0,0)[l]{$\vec{t'}$}}
\put(448,192){\makebox(0,0)[l]{$-\vec \nabla h(\vec 0)$}}
\put(566,293){\makebox(0,0)[l]{$-\vec \nabla h(\vec t)$}}
\sbox{\plotpoint}{\rule[-0.400pt]{0.800pt}{0.800pt}}%
\put(217,474){\vector(1,-1){239}}
\sbox{\plotpoint}{\rule[-0.200pt]{0.400pt}{0.400pt}}%
\multiput(342.58,585.55)(0.500,-0.612){425}{\rule{0.120pt}{0.590pt}}
\multiput(341.17,586.78)(214.000,-260.776){2}{\rule{0.400pt}{0.295pt}}
\put(556,326){\vector(3,-4){0}}
\sbox{\plotpoint}{\rule[-0.400pt]{0.800pt}{0.800pt}}%
\sbox{\plotpoint}{\rule[-0.200pt]{0.400pt}{0.400pt}}%
\multiput(217.00,474.58)(0.548,0.499){225}{\rule{0.539pt}{0.120pt}}
\multiput(217.00,473.17)(123.882,114.000){2}{\rule{0.269pt}{0.400pt}}
\put(342,588){\vector(1,1){0}}
\sbox{\plotpoint}{\rule[-0.400pt]{0.800pt}{0.800pt}}%
\sbox{\plotpoint}{\rule[-0.200pt]{0.400pt}{0.400pt}}%
\multiput(336.00,355.58)(0.554,0.499){201}{\rule{0.543pt}{0.120pt}}
\multiput(336.00,354.17)(111.873,102.000){2}{\rule{0.272pt}{0.400pt}}
\put(449,457){\vector(1,1){0}}
\put(528,720){\usebox{\plotpoint}}
\put(528.00,720.00){\usebox{\plotpoint}}
\put(510.45,708.96){\usebox{\plotpoint}}
\put(492.90,697.94){\usebox{\plotpoint}}
\put(475.59,686.55){\usebox{\plotpoint}}
\put(458.28,675.17){\usebox{\plotpoint}}
\put(441.46,663.08){\usebox{\plotpoint}}
\put(424.45,651.27){\usebox{\plotpoint}}
\put(407.72,639.04){\usebox{\plotpoint}}
\put(391.10,626.66){\usebox{\plotpoint}}
\put(374.70,613.96){\usebox{\plotpoint}}
\put(358.40,601.12){\usebox{\plotpoint}}
\put(342.24,588.24){\usebox{\plotpoint}}
\put(326.30,574.98){\usebox{\plotpoint}}
\put(310.53,561.53){\usebox{\plotpoint}}
\put(295.11,547.69){\usebox{\plotpoint}}
\put(279.74,533.79){\usebox{\plotpoint}}
\put(264.36,519.89){\usebox{\plotpoint}}
\put(248.99,505.99){\usebox{\plotpoint}}
\put(234.10,491.62){\usebox{\plotpoint}}
\put(219.25,477.20){\usebox{\plotpoint}}
\put(204.87,462.30){\usebox{\plotpoint}}
\put(190.45,447.45){\usebox{\plotpoint}}
\put(176.30,432.30){\usebox{\plotpoint}}
\put(162.60,416.75){\usebox{\plotpoint}}
\put(148.53,401.53){\usebox{\plotpoint}}
\put(134.81,386.01){\usebox{\plotpoint}}
\put(120.91,370.64){\usebox{\plotpoint}}
\put(108.12,354.40){\usebox{\plotpoint}}
\put(95.24,338.24){\usebox{\plotpoint}}
\put(88,330){\usebox{\plotpoint}}
\put(616,576){\usebox{\plotpoint}}
\put(616.00,576.00){\usebox{\plotpoint}}
\put(598.32,565.19){\usebox{\plotpoint}}
\put(581.10,553.66){\usebox{\plotpoint}}
\put(563.88,542.13){\usebox{\plotpoint}}
\put(546.66,530.59){\usebox{\plotpoint}}
\put(529.54,518.90){\usebox{\plotpoint}}
\put(513.06,506.44){\usebox{\plotpoint}}
\put(496.69,493.81){\usebox{\plotpoint}}
\put(479.87,481.70){\usebox{\plotpoint}}
\put(463.81,468.61){\usebox{\plotpoint}}
\put(447.40,456.05){\usebox{\plotpoint}}
\put(431.44,442.83){\usebox{\plotpoint}}
\put(415.89,429.17){\usebox{\plotpoint}}
\put(400.34,415.50){\usebox{\plotpoint}}
\put(384.81,401.81){\usebox{\plotpoint}}
\put(369.67,387.67){\usebox{\plotpoint}}
\put(355.06,373.06){\usebox{\plotpoint}}
\put(339.92,358.92){\usebox{\plotpoint}}
\put(325.32,344.32){\usebox{\plotpoint}}
\put(311.18,329.18){\usebox{\plotpoint}}
\put(296.64,314.51){\usebox{\plotpoint}}
\put(282.96,298.96){\usebox{\plotpoint}}
\put(269.31,283.41){\usebox{\plotpoint}}
\put(255.54,267.93){\usebox{\plotpoint}}
\put(242.42,251.90){\usebox{\plotpoint}}
\put(228.93,236.22){\usebox{\plotpoint}}
\put(220,225){\usebox{\plotpoint}}
\put(536,725){\usebox{\plotpoint}}
\put(536.00,725.00){\usebox{\plotpoint}}
\put(518.45,713.96){\usebox{\plotpoint}}
\put(500.90,702.94){\usebox{\plotpoint}}
\put(483.59,691.55){\usebox{\plotpoint}}
\put(466.28,680.17){\usebox{\plotpoint}}
\put(449.16,668.52){\usebox{\plotpoint}}
\put(432.05,656.84){\usebox{\plotpoint}}
\put(415.30,644.64){\usebox{\plotpoint}}
\put(398.97,631.97){\usebox{\plotpoint}}
\put(382.06,620.06){\usebox{\plotpoint}}
\put(366.04,607.04){\usebox{\plotpoint}}
\put(349.91,594.13){\usebox{\plotpoint}}
\put(333.84,581.10){\usebox{\plotpoint}}
\put(317.89,567.89){\usebox{\plotpoint}}
\put(302.27,554.36){\usebox{\plotpoint}}
\put(286.69,540.75){\usebox{\plotpoint}}
\put(271.32,526.85){\usebox{\plotpoint}}
\put(256.42,512.53){\usebox{\plotpoint}}
\put(241.33,498.33){\usebox{\plotpoint}}
\put(226.63,483.78){\usebox{\plotpoint}}
\put(211.56,469.56){\usebox{\plotpoint}}
\put(197.41,454.41){\usebox{\plotpoint}}
\put(182.88,439.60){\usebox{\plotpoint}}
\put(168.98,424.23){\usebox{\plotpoint}}
\put(154.97,408.97){\usebox{\plotpoint}}
\put(141.19,393.48){\usebox{\plotpoint}}
\put(127.29,378.11){\usebox{\plotpoint}}
\put(114.45,361.81){\usebox{\plotpoint}}
\put(100.63,346.38){\usebox{\plotpoint}}
\put(94,337){\usebox{\plotpoint}}
\put(624,580){\usebox{\plotpoint}}
\put(624.00,580.00){\usebox{\plotpoint}}
\put(606.11,569.66){\usebox{\plotpoint}}
\put(588.43,558.86){\usebox{\plotpoint}}
\put(571.21,547.33){\usebox{\plotpoint}}
\put(553.99,535.79){\usebox{\plotpoint}}
\put(536.85,524.13){\usebox{\plotpoint}}
\put(520.32,511.59){\usebox{\plotpoint}}
\put(503.53,499.43){\usebox{\plotpoint}}
\put(486.61,487.46){\usebox{\plotpoint}}
\put(470.52,474.52){\usebox{\plotpoint}}
\put(454.47,461.48){\usebox{\plotpoint}}
\put(438.43,448.43){\usebox{\plotpoint}}
\put(422.70,434.96){\usebox{\plotpoint}}
\put(406.75,421.75){\usebox{\plotpoint}}
\put(391.61,407.61){\usebox{\plotpoint}}
\put(376.13,393.85){\usebox{\plotpoint}}
\put(361.40,379.40){\usebox{\plotpoint}}
\put(346.26,365.26){\usebox{\plotpoint}}
\put(331.65,350.65){\usebox{\plotpoint}}
\put(317.51,335.51){\usebox{\plotpoint}}
\put(302.91,320.91){\usebox{\plotpoint}}
\put(289.25,305.34){\usebox{\plotpoint}}
\put(275.16,290.16){\usebox{\plotpoint}}
\put(261.55,274.55){\usebox{\plotpoint}}
\put(248.19,258.74){\usebox{\plotpoint}}
\put(234.87,242.87){\usebox{\plotpoint}}
\put(226,232){\usebox{\plotpoint}}
\end{picture}
\caption{Schematic drawing for the calculation of flow convergence.  The dotted curved lines are height contours.
\label{fig:convergence}}
\end{figure}
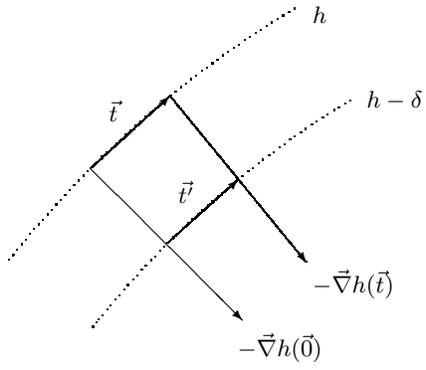

To begin, we define and calculate the convergence of flow on the surface.  Intuitively, a flow is convergent when neighboring flow lines approach each other as they continue downhill.  Consider the paths of two flow lines separated by an infinitesimally small distance $\vec t$ at height $h$.  A little further downhill, at a height $h-\delta$, they are separated by a distance $\vec{t'}$.  Figure~\ref{fig:convergence} illustrates the situation.  The rate of lateral approach is given by 
\begin{equation}
p= {|\vec t|-|\vec{t'}|\over |\vec t|\delta},
\label{notion}
\end{equation}
meaning that the convergence $p$ is the relative contraction of the contour line segment per unit height.  Dividing by $\delta$ makes the definition independent of the spacing between contours.

To obtain $p$, we first find the direction of the tangent to the contour, described by $\vec t= (x(t),y(t))$.  The parameter $t$ can be chosen to represent the length of the contour element.  We denote $h(t)\equiv h(x(t),y(t))$.  A first-order expansion
\begin{equation}
h(x,y)=h(0,0)+\partial_x h(0,0) x +\partial_y h(0,0) y  +O(x^2+y^2)
\end{equation}
yields the equation for the tangent $\vec t$ of the contour at point $(0,0)$.  Since $h(t)=\mbox{const.}$ and $\left|\vec t\right|=t$, one obtains to first order
\begin{equation}
x= t\frac{\partial_y h}{|\nabla h|}\quad\mbox{and}\quad
y=-t\frac{\partial_x h}{|\nabla h|}.
\label{tangent}
\end{equation}
Derivatives without their argument are here always understood to be taken at the origin, that is, we write $\partial_x h$ in lieu of $\partial_x h(0,0)$, and so on.  The first-order expansion of the gradient along the contour becomes
\begin{equation}
\nabla h(t)=\nabla h(0)+{t\over|\nabla h(0)|}\left(
	\begin{array}{c} 
		 \partial_{xx}h \partial_y h - \partial_{xy}h\partial_x h\\ 
                 \partial_{xy}h \partial_y h - \partial_{yy}h\partial_x h
        \end{array}
\right) +O(t^2).
\label{gradexpansion}
\end{equation}
According to Fig.~\ref{fig:convergence},
\begin{eqnarray}
\vec{t'}&=&\vec t+{\delta\over|\nabla h(t)|}{\nabla h(t)\over|\nabla h(t)|}
               -{\delta\over|\nabla h(0)|}{\nabla h(0)\over|\nabla h(0)|} \nonumber\\
        &=&\vec t+{\delta\over|\nabla h(0)|^2}\left[\nabla h(t)-\nabla h(0)\right]+O(t \delta).
\label{eq:tmp}
\end{eqnarray}
A straight-forward calculation using eqs.~(\ref{notion}), (\ref{gradexpansion}), and (\ref{eq:tmp}) then yields 
\begin{equation}
p={(\partial_y h)^2\partial_{xx}h -2(\partial_x h)(\partial_y h)\partial_{xy}h +(\partial_x h)^2\partial_{yy}h
\over \left[(\partial_xh)^2+(\partial_yh)^2\right]^2}.
\label{convergence}
\end{equation}
The surface is convergent when the expression (\ref{convergence}) is positive and divergent when it is negative.  The expression for the curvature of the contour differs from eq.~(\ref{convergence}) only by a denominator of $|\nabla h|^3$ \cite{bronshtein} instead of $|\nabla h|^4$, so that the convergence may also be written as the curvature of the contour, $\kappa$, divided by the slope, $p=\kappa/|\nabla h|$.

The definition of convergence (\ref{notion}) is merely based on the geometry of the flow paths and does not involve velocities.  In fact, without further assumptions one has no knowledge of velocities and fluxes on the surface.  

Essentially the same notion of convergence as in eq.~(\ref{notion}) has been proposed earlier \cite{dit93} and employed in studies of landscapes \cite{dit93,mfg93}.  In ref.~\cite{dit93} landscapes were decomposed into elements by tracing their height contours.  The generated contour segments were then used to evaluate $(|\vec t|-|\vec{t'}|)/(|\vec t|+|\vec{t'}|)$ for a given $\delta$.  For an infinitesimally fine grid this agrees, up to a prefactor, with eqs.~(\ref{notion}) and (\ref{convergence}).  Others (see refs.~\cite{picturebook,rinaldo95}) have proposed the Laplacian $\nabla^2h$ as a local definition of convergence, which can differ from eq.~(\ref{convergence}) even in its sign.  The use of the Laplacian applies for a particular model of fluxes and erosion, but it is not a general expression for topographic convergence.  For example, the hyperbolic surface $h(x,y)=2x^2-y^2$ would be erroneously identified as convergent everywhere.  The proper distinction between topographic convergence and divergence is given by the sign of eq.~(\ref{convergence}).

\begin{figure}
\noindent
a)\epsfig{file=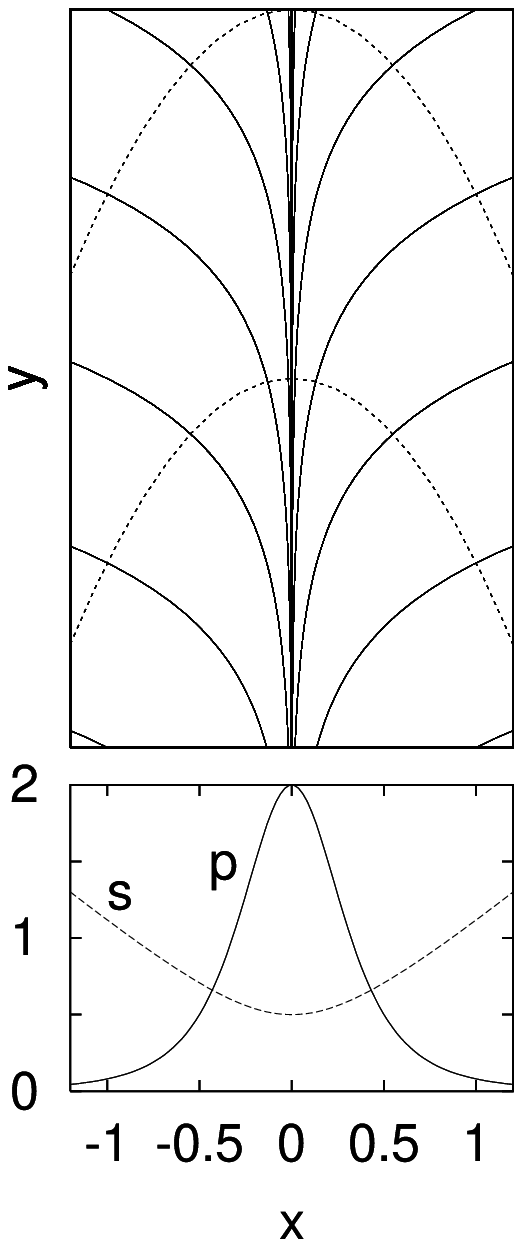,width=3.9cm}
b)\epsfig{file=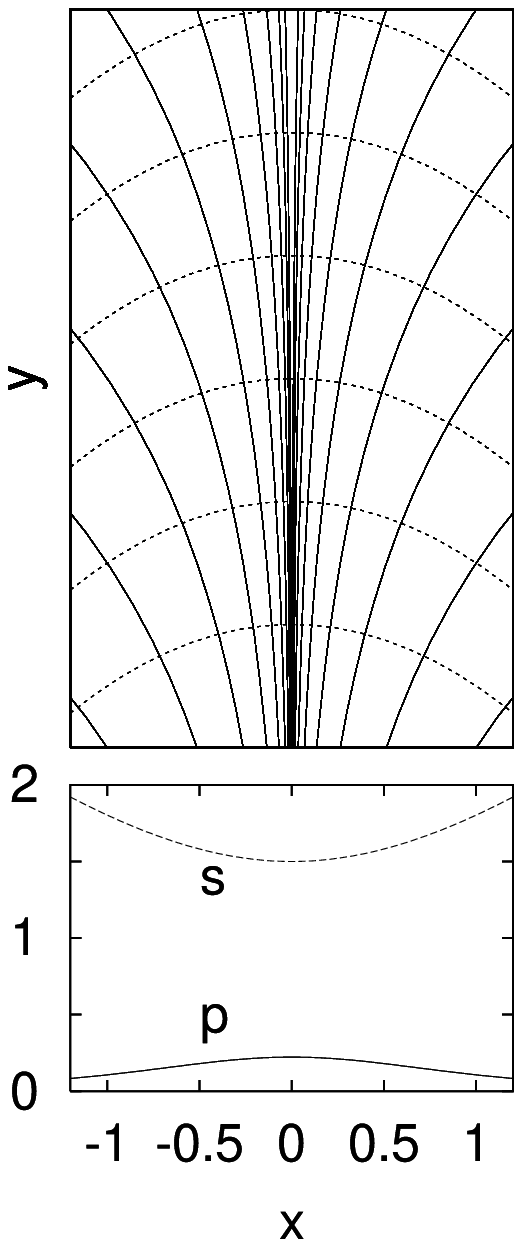,width=3.9cm}
\caption{Illustration of the effect of slope on the convergence of flow lines.  A simple surface of the form $h(x,y)=x^2+by$ is used as an easily visualized example.  The solid lines in the upper panels of the figure show the flow lines for (a) $b=1$ and (b) $b=3$.  The dotted lines are height contours.  The contour interval is the same in both parts of the figure.  One sees the stronger convergence in the flatter of the two surfaces, (a), as well as the stronger convergence in the flatter regions towards the center of the x-axis within each figure.  The behavior of the convergence $p(x)$ and the slope $s(x)$ (actually $s(x)/2$) are illustrated in the lower panels. 
\label{fig:anotherill}}
\end{figure}

It is worthwhile to pause here to contemplate the intuitive reasons for convergence.  If the contours are strongly curved the surface acts like a funnel and water converges rapidly.  The slope influences the convergence via the curvature, but it also alters the available path length per unit height.  The latter effect is reflected in the denominator of $p=\kappa/|\nabla h|$.  Fig.~\ref{fig:anotherill} demonstrates this slope effect in a simple example.  In this example, both $\kappa$ and $|\nabla h|$ change to increase the convergence.  On steep slopes water falls quickly along paths with little relative motion, while on flatter slopes paths are deflected more easily.

The notion of convergence intrinsically requires that the elevation is a function of two spatial coordinates.  It has no equivalent in a one-dimensional setting, where the height is considered as a function of one coordinate only.  Hence, any consequences derived from the convergence expression are three-dimensional in nature (height and two horizontal coordinates).

In our idealized setting rivers may be associated with locations carrying much more running water than neighboring points at the same height.  They can do this by accumulating water from uphill more effectively.  If a point converges flow faster than a neighboring point at the same height, fluid moves towards the point of higher convergence.  Consequently, locations of strongest convergence tend to act as rivers.  It is thus clear that rivers tend to form in locations of large $p$, where the slope is low and the curvature high.  Rivers are thus associated with a local optimality property, resulting from the mere fact that they flow on surfaces.  This optimality property is not directly equivalent to ``optimal channel networks'' \cite{OCN_review} or any of the other proposed optimality principles \cite{leopold} for river networks.

\section{Gaussian Surfaces}
\label{sec:gaussian} 

Gaussian surfaces are random continuous surfaces described by a superposition of waves with random phases \cite{adler,lahi57a,lahi57b,isi92}.  Results on Gaussian surfaces have been applied in various fields, including electric noise \cite{rice54}, water waves \cite{lahi57a,lahi57b}, beaches \cite{hoho93}, image processing \cite{oak98}, and the clustering of galaxies \cite{bar86}.  We will here need only some of their elementary properties.

A Gaussian surface is defined as 
\begin{equation}
h(x,y)=\sum_{k,\ell} c(k,\ell) e^{-i(kx+\ell y+\varphi(k,\ell))}
\label{gsurface}
\end{equation}
where $\varphi$ is a random variable uniformly distributed between 0 and $2\pi$.  The amplitudes $c(k,\ell)$ are arbitrary, but, of course, they need to decay sufficiently fast with wavenumber to guarantee the continuity and differentiability of the surface.  Any Gaussian surface has the property that its heights and all of its derivatives are distributed Gaussian.  Furthermore, the first derivative is statistically independent of the height, and the second derivatives are independent of the first derivatives \cite{adler}.  

This last property allows us to evaluate the convergence for a given slope quantitatively.  Qualitatively, it is already clear from our early discussion of expression~(\ref{convergence}) that flatter areas tend to converge flow more rapidly.  For Gaussian surfaces this relation can be made precise.  The convergence, $p$, averaged over all points of a given slope, $|\nabla h|$, is called the conditional average of the convergence on the slope, $\av{p| \, |\nabla h|}$.  The simple conditional average $\av{p | (\nabla h)^2}=0$ vanishes, because $p$ changes sign as $h\to -h$.  Thus we look at the square of the convergence:
\begin{eqnarray*}
\av{p^2| (\nabla h)^2} &=&
	\av{(\partial_{xx}h)^2}\av{\left.{(\partial_y h)^4\over (\nabla h)^8}\right| (\nabla h)^2} + 
	\\
&&	+\av{(\partial_{yy}h)^2}\av{\left.{(\partial_x h)^4\over (\nabla h)^8}\right| (\nabla h)^2} + 
	\\
&&	+4\av{(\partial_{xy}h)^2}\av{\left.{(\partial_x h)^2(\partial_y h)^2\over (\nabla h)^8}\right| (\nabla h)^2} +
	\\
&&	+2\av{(\partial_{xx}h)\partial_{yy}h}\av{\left.{(\partial_x h)^2(\partial_y h)^2\over (\nabla h)^8}\right| (\nabla h)^2}
\end{eqnarray*}
Here, we have made use of the statistical independence of first and second derivatives.  The mixed terms of $p^2$ vanish because of the statistical symmetries $x\to -x$ and $y\to -y$.  The conditional averages now reduce to geometrical factors, which can be rewritten for a statistically isotropic situation in terms of an angle $\theta=\tan^{-1}(\partial_yh/\partial_xh)$:
\begin{widetext}
\begin{eqnarray}
\av{p^2| (\nabla h)^2} &=&
 	\av{(\partial_{xx}h)^2}\av{\sin^4\theta}{1\over (\nabla h)^4} + 
	\av{(\partial_{yy}h)^2}\av{\cos^4\theta}{1\over (\nabla h)^4} + \nonumber\\
&&      +4\av{(\partial_{xy}h)^2}\av{\sin^2\theta \cos^2\theta} {1\over (\nabla h)^4} + \nonumber\\
&& 	+2\av{(\partial_{xx}h)\partial_{yy}h}\av{\sin^2\theta \cos^2\theta} {1\over (\nabla h)^4}\\
&=& 	\left[
	\frac{3}{8}\av{(\partial_{xx}h)^2}+ 
	\frac{5}{2}\av{(\partial_{xy}h)^2}+
	\frac{5}{4}\av{(\partial_{xx}h)\partial_{yy}h}+
	\frac{3}{8}\av{(\partial_{yy}h)^2} \right]
		{1\over (\nabla h)^4} \label{cg1}\\
&\propto& 1/(\nabla h)^4 \label{cg2} 
\end{eqnarray}
\end{widetext}
Therefore, the convergence behaves as the inverse of slope squared, 
\begin{equation}
|p|\sim 1/(\nabla h)^2
\label{cg3}
\end{equation}
or vice versa, $|\nabla h|\sim1/\sqrt{|p|}$.  Since the slope enters quadratically in (\ref{cg3}), it has a {\em strong} effect on the convergence.  This is the relation between convergence and slope for a Gaussian surface, irrespective of the surface's spatial correlations.  In fact, for result~(\ref{cg1}) to be valid, the surface does not necessarily need to be Gaussian.  Instead, one requires only that the first and second derivatives of height be statistically independent of each other. 

We can check this result on both synthetic and real landscapes.  A Gaussian surface can be efficiently created using Fourier transforms \cite{footnote1}.  Interpolation could be used to obtain heights and gradients at any position between grid points.  Real landscapes from digital elevation maps, which we use below, are however conventionally evaluated on a discrete grid.  We opt to evaluate synthetic surfaces and real landscapes in the same discrete way.  Fluid paths are determined by following the direction of steepest descent among the eight neighbors.  Fluid trajectories are stopped when they reach a local minimum, implying a lake or ocean at that place.  We choose the smallest wavelength $\lambda$ as four grid cells long, which is twice the smallest representable wavelength.  The largest wavelength $\Lambda$ equals the domain size.  

\begin{figure}
\epsfig{file=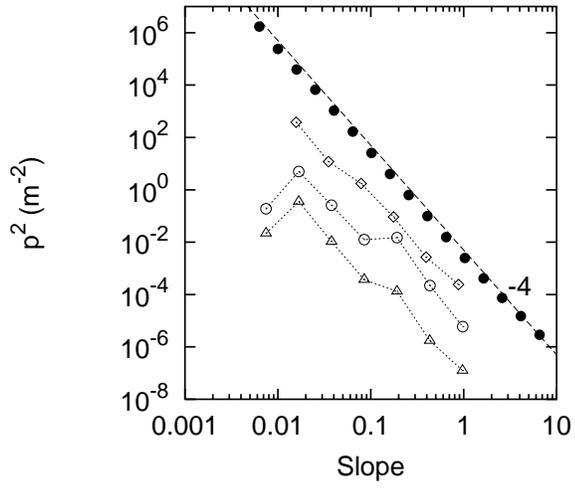,width=8cm}
\caption{Convergence-slope relation for synthetic (filled dots) and real landscapes (unfilled symbols).  The dashed line shows the analytical result for the synthetic landscape.  All data are averaged over logarithmically spaced bins on the abscissa.  The unfilled symbols correspond to three different landscapes in Northern California as described in the text. (The units for the synthetic landscape are arbitrary and hence only the slope of the graph can be compared to real data.  The two lowest graphs are shifted downwards by factors of 40 and 1600, respectively, for better visibility.)
\label{fig:con_cg}}
\end{figure}

Figure~\ref{fig:con_cg} shows the convergence-slope relation measured in the simulated surfaces, which must agree with formula (\ref{cg1}).  The filled dots are the measurements from a simulated surface and the dashed line is the theoretical calculation using the prefactor in eq.~(\ref{cg1}).  The tiny deviations between theory and numerics have two sources:  The average was only taken over one realization and the domain is periodic in both directions.  In the same figure are data from real landscapes, discussed later in the text.  Note now, however, the good fit over two orders of magnitude in slope; almost the entire range.

The result~(\ref{cg1}) shown in Fig.~\ref{fig:con_cg} is independent of the choice for the power spectrum.  Yet, for completeness, we note that the spectrum used in this example is a power-law spectrum, where the coefficients in eq.~(\ref{gsurface}) are $|c(\vec k)|\propto 1/|\vec k|^{(1+\rho)}$ and the phases are random.  The exponent $\rho$ is known as the Hurst exponent.  It has a simple relation with the ``roughness exponent'' that characterizes the spatial correlation of the height.  The structure function $\av{|h(x)-h(x+R)|^2}$ is proportional to $R^{2\rho}$ for $0<\rho<1$.  If $\rho$ is small, one has to introduce a cut-off scale to maintain the differentiability of the surface, and therefore we take $|c(\vec k)|=1/|\vec k|^{(1+\rho)}$ for $2\pi/\lambda<k<2\pi/\Lambda$ and $c(\vec k)=0$ otherwise.  Graphs with solid dots, in Fig.~\ref{fig:con_cg} and elsewhere, correspond to a Hurst exponent of $\rho=0.7$.  

In concluding this section, we note that convergence is intuitively related to the drainage area.  Points of stronger convergence will, on average, possess larger drainage areas, because they collect more flow than points of low convergence.  If $p$ increases so does $a$, leading to a slope-area relation.

\section{Slope-Area Relation}

Erosion rates increase with the amount of flowing water and are also affected by the local slope of the landscape, among other factors \cite{sch61}.  If other parameters, like width, depth, and velocity of a river are considered a function of total discharge and slope, one arrives at the conventional assumption that erosion is a function of slope and drainage area only \cite{picturebook}.  Local topography can carry the signature of the entire basin draining into it \cite{chan00}.

\begin{figure}
\noindent
a)\epsfig{file=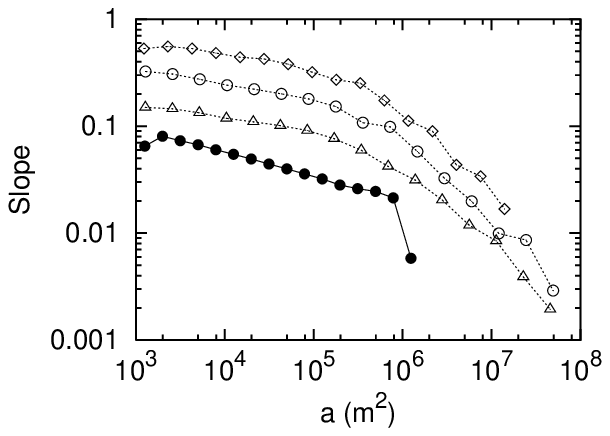,width=8cm}
b)\epsfig{file=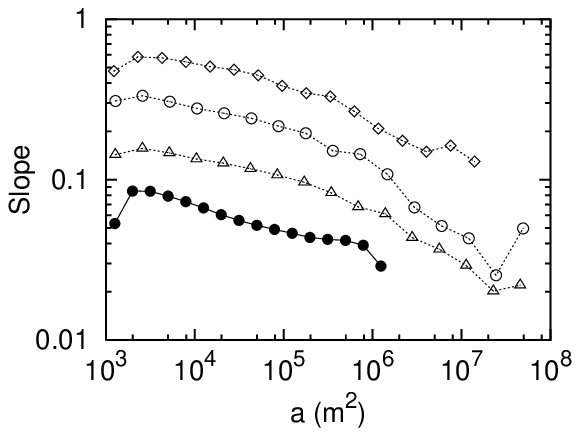,width=8cm}
\caption{Slope-area relation for synthetic and real landscapes.  The solid line with filled dots in parts (a) and (b) is for a simulated Gaussian surface at a resolution of $4096\times 4096$.  The length units for the simulated landscape are chosen arbitrarily.  The unfilled symbols correspond to Juan Creek ($\diamond$), Middle fork of the Noyo river ($\circ$), and the North fork of the Noyo river ($\bigtriangleup$).  These symbols are used uniformly in all figures.  (The graphs for the last two landscapes are shifted downwards by factors of 1.4 and 3, respectively, for better visibility.)  The two parts of the figure differ only in the way the slope is determined.  In both parts real and artificial landscapes exhibit similar slope-area graphs, suggesting the relation has a simple geometric origin. 
\label{fig:con_ga}}
\end{figure}

On real landscapes a statistical relation is observed between the local slope and the drainage area $a$.  Its quantitative form is roughly $|\nabla h| \sim 1/\sqrt{a}$ \cite{hk83,tbr89,picturebook}.  There are however substantial variations of the exponent of 0.5 for different river systems \cite{will91}.  
We have analyzed real landscapes from digital elevation maps of coastal landscapes from the Mendocino region in northern California \cite{snyder00,sd98}.  A detailed description of the Juan Creek data is given in ref.~\cite{snyder00} and the Noyo river basin is discussed in ref.~\cite{sd98}.  These data have 30m resolution in both horizontal directions.  Fig.~\ref{fig:con_ga}a shows the slope-area relation of a Gaussian surface together with that of three  real landscapes.  One sees that the random surface exhibits a slope-area relation similar to the one seen in the real data \cite{footnote2}.

For Fig.~\ref{fig:con_ga}a the slope was calculated as the maximum drop from pixel to pixel.  Another method, used in all other figures, is to take a finite difference in both horizontal directions and to evaluate the absolute value of the gradient in this way.  Using a stencil of five points in both direction, this is two orders more accurate than the previous method, but not necessarily more realistic, because the gradient is in this case not exactly aligned with the direction of the flow.  Figure~\ref{fig:con_ga}b repeats Fig.~\ref{fig:con_ga}a, with the alternative slope measurement.  The observed slope-area relationships are weaker in this case, but still roughly equivalent for real and Gaussian surfaces.  The discrepancy with Fig.~\ref{fig:con_ga}a arises merely from the ambiguity of measuring the slopes from the data \cite{footnote3}.

Real landscapes are more complex than Gaussian surfaces.  Erosion might alter landscapes in a way not captured by Gaussian surfaces.  Yet, the direct comparisons of the slope-area relation in Fig.~\ref{fig:con_ga} shows a similarity between them.  The extent to which the geometric explanation of the slope-area relation applies to real landscapes can be further probed by looking at the convergence-slope and convergence-area relation of the real landscapes.  Fig.~\ref{fig:con_cg} shows that the convergence-slope relation of Gaussian surfaces is reasonably true also for real landscapes.  All of them show a behavior close to $p^2\sim 1/|\nabla h|^4$.  Fig.~\ref{fig:con_ca} shows the convergence-area relation for real landscapes together with that for a synthetic landscape.  Particularly small drainage areas are found for divergent points.  The minimum drainage is the size of one cell, which is $900\mbox{m}^2$.  After a rapid initial growth of the average convergence for smaller areas, the graph approaches its asymptotic behavior.  Again, we see that real landscapes behave in this respect similar to Gaussian surfaces.  Hence, there is no doubt that the geometric effect found for Gaussian landscapes is also present in real landscapes.

\section{Variations in the Slope-Area Relation on Gaussian Surfaces}
\label{sec:new}

The slope-area relation depends on the statistical parameters of the surface, unlike eq.~(\ref{cg1}).  Therefore, we obtain pertinent statistics from numerical simulations.  In the following, we discuss how the relation changes with the choice of the power spectrum.  We explore various surfaces that vary over a wide range of length-scales.

\begin{figure}
\epsfig{file=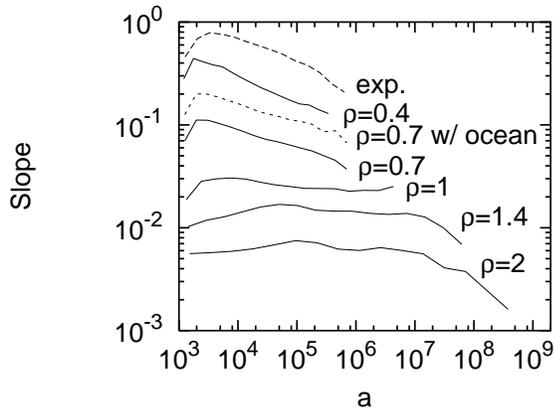,width=8cm}
\caption{Slope-area relation for Gaussian landscapes.  The different graphs correspond to landscapes with power-law spectra with different Hurst exponents, an exponential power spectrum, and a landscape half-filled with an ``ocean''.  They are obtained at a resolution of $1024\times 1024$.  The length units are chosen arbitrarily.
\label{fig:con_ga-2}}
\end{figure}

Fig.~\ref{fig:con_ga-2} shows the slope-area relation for power laws with several different Hurst exponents $\rho$, one exponential power spectrum, and a landscape that is half-filled with water (ignoring the lower half when taking the statistics).  For the surfaces with short spatial correlation there is a clear area dependence of the slope, while the surfaces that change more slowly exhibit less of a dependence.  For landscapes that are statistically self-similar, and hence characterized by their Hurst exponent $\rho$, the slope-area relationship becomes more and more pronounced as $\rho$ goes from one to zero.

Real landscapes are said to have Hurst exponents of about 0.7 \cite{mandelbrot}, although there are considerable variations among individual landscapes (see \cite{turcotte} and refs. in \cite{roth98}).  We determined the structure functions of the real landscapes used in this study.  They exhibit only a rather limited range of scaling, if at all, but they tend to be substantially flatter than 0.7.  This corresponds to a lower Hurst exponent and thus a faster decorrelation.  However, given the average tilting of the coastal basins and their small size, this might not be significant.  In any case, in the range of Hurst exponents $0<\rho<1$ the simulated Gaussian surfaces show a clear slope-area dependence, as seen in Fig.~\ref{fig:con_ga-2}.  The graphs for $\rho=0.4$ and $\rho=0.7$ have a slope in the range of 0.15--0.25.  Varying the power spectrum can lead to flatter or steeper slopes, but we have never observed a slope-area graph steeper than about 0.3.

\begin{figure}
a)\epsfig{file=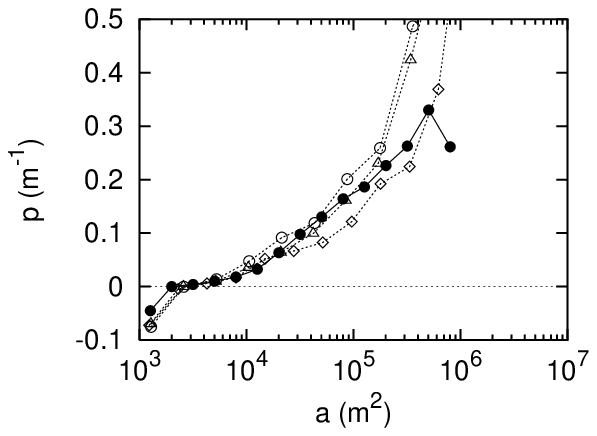,width=8cm}
b)\epsfig{file=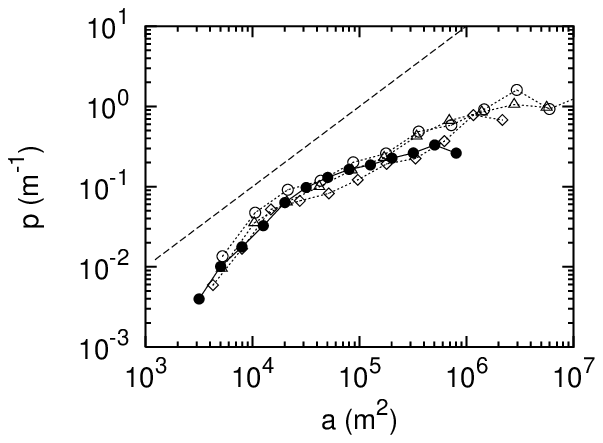,width=8cm}
\caption{Convergence-area relation for synthetic (filled symbols) and real landscapes (empty symbols).  The two plots show the same data, $\av{p|a}$, on (a) a semi-logarithmic and (b) a log-log plot. The length units for the simulated landscape are chosen arbitrarily.   The dashed line in (b) corresponds to $p\propto a$.  
\label{fig:con_ca}}
\end{figure}

Fig.~\ref{fig:con_ca}b shows that for large convergence, $p$ grows slower than the area $a$ for the real and the simulated landscapes.  The same is true for simulated landscapes with other parameters (not shown in Fig.~\ref{fig:con_ca}) and we have not found any Gaussian landscape that would violate this trend.  Since $p\le O(a)$ and $p\sim 1/|\nabla h|^2$, it follows that $|\nabla h|\ge O(1/\sqrt{a})$.  Hence, one expects that slope decays no faster than $1/\sqrt{a}$.  As mentioned above, the actual restriction appears to be even stronger, around $a^{-0.3}$.  This indicates the presence of a bound for the area-dependence of the slope on a Gaussian landscape.  The convergence depends strongly on the slope and the area depends strongly on the convergence, and therefore the area strongly on the slope, or, equivalently, the slope weakly on the area.  Hence, it is difficult to get a particularly strong slope-area relation through geometrical effects alone.

In the particular case of Fig.~\ref{fig:con_ga} the graphs for the real landscapes change their slope at about $10^6\mbox{m}^2$.  This corresponds geologically to a transition from colluvial (loose debris) channels to bedrock channels \cite{snyder00}.  The fact that the simulations nowhere exhibit the final drop in the slope-area relation to its full extent seen in real landscapes (Fig.~\ref{fig:con_ga}), indicates that there are contributions to the slope-area relation beyond the geometrical effect we have unearthed in this paper.  Exceptional cases of real landscapes are known with slope-area relations that are substantially more pronounced than the ones we encounter on Gaussian surfaces \cite{sd98,kw00}.  This leaves room for erosion, tectonics, or further mechanisms contributing to the slope-area relation.  However, since a substantial slope-area relation is produced by Gaussian surfaces with realistic Hurst exponents, the slope-area relation carries less information about these processes than previously thought.

\section{Discussion}
\label{sec:end}

In this work, we found and explained a correlation between slope and drainage-basin area on random topography.  We gave a two-step argument.  The first step was to derive analytically the convergence expression and to show explicitly the dependence of convergence on slope.  The second step connected the convergence with the area of drainage.  This connection is intuitive, but we also demonstrated it numerically.  Further, we have left little doubt that the same effect appears on real landscapes too, by verifying the validity of both steps with real data.  

We have also found that the slope-area correlation cannot be strengthened indefinitely by varying the parameters.  The slope-area relation is often approximated as $|\nabla h| \sim a^{-\theta}$.  Since we find for random surfaces $0 \le \theta \le 0.3$, an exponent in this range implies nothing about geological processes.  It is instead only a consequence of water flowing downhill on random landscapes.  This is in direct contrast to hypotheses that attribute any slope-area relation to erosional mechanisms (see refs. in \cite{picturebook}).  

If erosional mechanisms were responsible for slope-area relations one would expect a deterministic relation between slope and area.  A certain drainage area would always lead to the same slope at its outlet.  Fluctuations would then be due to variations in geological parameters.  On Gaussian surfaces on the other hand there are intrinsic statistical fluctuations and the slope-area relation is a statistical correlation rather than a deterministic formula.  Investigations of qualitative differences between slope-area fluctuations on real and random landscapes are therefore expected to yield additional insight.

Although this study is motivated by geomorphology the same statistical findings are applicable to other physical systems, where the motion is strictly along the gradient of the force field and the inertia negligible.  The slope-area relation in this context states that for a quickly decorrelating force field the domain of attraction will, on average, be larger when the force is smaller.  (Ref.~\cite{isi92} reviews applications of Gaussian surfaces when the motion is along equi-potential lines, rather than perpendicular to them.)

The slope-are relation is but one consequence of the convergence expression~(\ref{convergence}).  Another consequence, as we have briefly discussed, is the formation of rivers preferably on locations of high convergence.  The local optimum is with respect to points on the same contour line, that is, $p$ is maximum along $h=\mbox{const.}$, or in terms of our earlier notation $dp/dt=0$ and $d^2p/dt^2<0$.  This indicates a previously overlooked optimality principle for the organization of rivers.  (Note that this extremal principle is not strictly valid for points at different heights.  If it were, it would immediately imply that the longitudinal stream profiles are concave).  Whether such an observation may by of any use is unclear.  What is clear, however, is that Gaussian or otherwise random topography has far-reaching non-trivial implications on basin statistics.

\begin{acknowledgments}
It is our pleasure to thank Daniel Collins, Peter Dodds, Bob Fleming, Eric Kerby, Noah Snyder, and Kelin Whipple for valuable discussions.  This work was supported by Department of Energy grant DE FG02-99ER 15004.
\end{acknowledgments}

%\bibliographystyle{revtex}
%\bibliography{/home/norbert/Papers/my,/home/norbert/Papers/geo} 

\begin{thebibliography}{10}
\bibitem{footnote1}
\bibinfo{note}{The spectral representation of the particular Fast Fourier
  Transform routine we used \cite{nrc} consisted of the real and imaginary
  value for the complex amplitude of each mode, which are assigned the
  amplitudes from the chosen power spectrum and random phases. The spectral
  representation has redundant elements \cite{nrc} and the symmetries in the
  Fourier transforms of a real field need to be imposed upon them. The Gaussian
  nature of height, gradients, statistical isotropy, and the aforementioned
  independence of derivatives are verified in the simulations.}
\bibitem{footnote2}
\bibinfo{note}{Incidentally, simulations in one dimension, where the ``area" is
  the distance to the adjacent maximum, do not show an inverse relation between
  area and slope over such a wide range of areas. This confirms that the
  slope-area relation originates from an effect involving both horizontal
  coordinates.}
\bibitem{footnote3}
\bibinfo{note}{In this paper averages are always taken over the area.
  Alternatively, one can consider an average along the main stream of the
  drainage basin. Preliminary evidence from the simulation of Gaussian surfaces
  indicates that the slope-area relation is in this case also fully present.}
\end{thebibliography}

\newpage

\end{document}